\documentclass[twocolumn,preprintnumbers,amsmath,amssymb]{revtex4}
\usepackage{graphicx}% Include figure files
\usepackage{dcolumn}% Align table columns on decimal point
\usepackage{bm}% bold math

\newcommand{\be}{\begin{equation}}
\newcommand{\bel}[1]{\begin{equation}\label{#1}}
\newcommand{\ee}{\end{equation}}
\newcommand{\bea}{\begin{eqnarray}}
\newcommand{\ba}{\begin{array}}
\newcommand{\eea}{\end{eqnarray}}
\newcommand{\ea}{\end{array}}

\begin{document}

\title{\bf Temporal and structural characteristics of a two dimensional gas of hard needles }

\author{M. Ebrahim Foulaadvand $^{1,2}$ \thanks{Corresponding Author: e-mail: foolad@iasbs.ac.ir,} and Mohsen Yarifard $^{1}$}

\affiliation{ $^1$ Department of Physics, Zanjan University, P.O.
Box 19839-313, Zanjan, Iran.}

\affiliation{$^2$ Computational Physical Sciences Laboratory,
Department of Nano-Science, Institute for Research in Fundamental
Sciences (IPM), P.O. Box 19395-5531, Tehran, Iran. }

\date{\today}
\begin{abstract}

We have simulated the dynamics of a 2D gas of hard needles by
event-oriented molecular dynamics. Various quantities namely
translational and rotational diffusion constants and intermediate
self scattering function have been explored and their dependence
on density is obtained. Despite absence of positional ordering,
the rotational degree of freedom behaves nontrivially. Slowing
down is observed in the angular part of the motion. It is shown
that above a certain density the rotational mean squared
displacement exhibits a three stage regime including a plateau.

\end{abstract}

\maketitle
\section{{Introduction}}

Molecular dynamics (MD) simulation of anisotropic hard objects
such as ellipsoids and spherocylinders has been the subject of
exploration in past decades \cite{allen89}. It has been shown the
alignment of non-spherical molecules can lead to a diversity of
phases, mainly orientational in nature, in liquid crystals
\cite{allen89,degennes,frenkel87}. Despite the profound insight
obtained via tremendous Monte Carlo (MC) simulations of the
static phases, many dynamical, transportational and structural
properties namely kinetic arrest and glassy behaviours have only
been poorly understood \cite{tanja}. In spite of employment of
other simulation techniques like Brownian dynamics
\cite{yamamoto,fixman1,tao} and theoretical approaches such as
kinetic theory \cite{huthmann,green}, density functional
\cite{mcgrother,vink,chrzanowska05} and hydrodynamics equations
approach \cite{otto}, event-oriented MD remains as an efficient
tool for probing the dynamical aspects of hard gases of
non-spherical objects. Among elongated and anisotropic hard
bodies, infinitely thin needles have received quite notable
attention \cite{allen89}. The first MD simulation attempt were
carried out for a three dimensional gas of hard needles by
Frenkel and Maguire \cite{frenkel81,frenkel83} and Magda et al.
\cite{magda}. In these investigations, various temporal
auto-correlations were explored and compared to predictions of
Enskog and Doi-Edward theories. The physics of hard needle gas is
sensitive to dimensionality. By extensive MC simulations in 2D,
Frenkel and Eppenga showed the existence of quasi long range
order in the system at high densities \cite{frenkel85}. It was
argued that this 2D system undergoes a Kosterlitz-Thouless phase
transition \cite{kosterlitz}. Resurgence of interest in the
dynamical properties of hard needle gas was sparked by papers of
Renner et al. \cite{renner} and Obukhov et al. \cite{obukhov} who
introduced a rotator model. Their simple ideal glass former can
mimic the basic dynamical properties of an orientational glass.
This rotator model was shown to exhibit orientational glassy
behaviour like its positional counterpart i.e., hard-sphere
system \cite{woodcock1}. A natural question to ask is whether the
orientational glassy behaviour survives if one releases the
positional degrees of freedom. Chrzanowska et al. carried out the
first MD simulation of a two dimensional hard needle gas
\cite{chrzanowska02,chrzanowska04} and mainly studied velocity
auto correlations whereas the transport properties remained
unexplored. Subsequently transport coefficients like
self-diffusion and shear viscosity of a 3D hard needle gas were
studied within MD approach by Muk\^{o}yama et al.
\cite{mukoyama1,mukoyama2}. We wish to note that the hard needle
rotator model is amenable to comparison with experiments carried
out via neutron scattering and has shown to successfully model
the experimentally observed phenomenon of orientation glassy
dynamics \cite{ruiz}. Recent investigations on 2D hard needle gas
involve deposition-evaporation dynamics \cite{barma} and
restriction of needles center of masses \cite{kardar}. Here we
focus on the transportation features to illuminate the interplay
of translational and rotational degrees of freedom in 2D systems
of highly anisotropic hard objects.

\section{ Description of the Problem }

Consider a system comprising of infinitely thin hard needles with
mass $m$ and length $l$ which are restricted to move in the two
dimensional $x-y$ plane. The interaction between needles is
assumed to be hard core. We work in the reduced units in which
$m$ and $l$ are taken unity. Upon a collision between two
needles, the centre of mass (CM) velocity and the angular
velocity around the $z$ axis through the CM will change. We
assume the collisions are elastic and frictionless therefore the
energy (entirely kinetic) is conserved after a collision. The
kinetics of such collision has been investigated in detail in
\cite{chrzanowska04}. We study the dynamics of this 2D needle gas
within the collision-oriented MD approach. In the
collision-oriented MD the evolution of system takes place from
collision to collision. The main task in MD of hard objects is to
find the collision time between two apart objects
\cite{raghavan}. A numerical scheme has been originally
introduced in \cite{sando,talbot}. The method is based on finding
a geometrical condition for overlapping between objects. This
condition is given as $H_1, H_2<\frac{l}{2}|\sin(\theta)|$ where
$\theta$ is the angle between needles. $H_2$ is the distance of
needle $2$ CM to the line of needle $1$ and reverse. To find the
next collision time we move all the needles in small time steps
until the needles become very close to each other that can be
regarded as touching. Then we update the velocity and angular
velocity of the colliding needles and proceed to the next
collision event.

\section{ Molecular dynamics simulation }

\subsection{ Static properties and velocity autocorrelations}

We have simulated the dynamics of a 2D gas of hard needles with
the method explained in section II. Periodic boundary condition is
imposed. The size of our simulation box is set to $L=7$. The
number density is defined as $\rho=\frac{N}{L^2}$ where $N$ is
the number of needles. The total energy of the system $E$
(entirely kinetic) is divided into two segments of translational
$E_{trs}$ and rotational $E_{rot}$. We remark that due to lack of
any energy scale in the potential energy between particles, the
temperature $T$ appears as an overall multiplicative factor in
thermodynamic quantities such as pressure and free energy. Thus
the state of the system trivially depends on temperature and
hence energy. Nevertheless $E$ determines the time scale $\tau$.
In the thermal unit one has $\tau=l\sqrt{\frac{m}{k_BT}}$. We have
taken the energy per particle $\frac{E}{N}=1.5$ in our
simulations which gives $k_BT=1$. A useful quantity is the
average number of collision per unit time $\Gamma$ that a needle
experiences. In \cite{frenkel83} and \cite{magda} it is argued
that $\Gamma$ has a linear dependence on number density $\rho$ in
a 3D gas. According to Frenkel and Maguire arguments in 3D,
$\Gamma$ scales as follows: $\Gamma \sim
l^2(\frac{kT}{m})^{\frac{1}{2}}\rho$. In fact the Doi-Edward
theory predicts such linear behaviour as well. We have computed
the dependence of $~\Gamma$ on $\rho$ in our 2D model. Figure (1)
depicts that $\Gamma$ has, analogous to the 3D case studied in
\cite{frenkel83,magda}, a linear dependence on $\rho$. Note the
slope change about $\rho=5.5$.

\begin{figure}
\centering
\includegraphics[width=7.5cm]{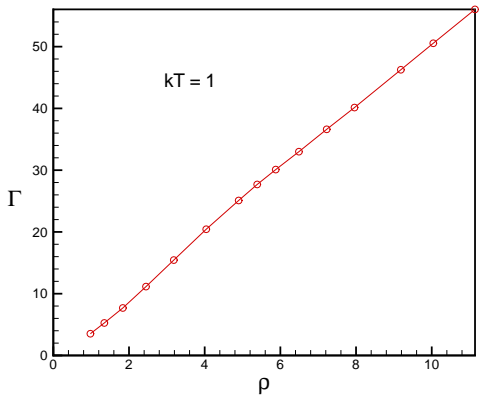}
\caption{ Fig.1: Collision frequency $\Gamma$ vs $\rho$. The
dependence is linear like 3D gas of hard needles investigated in
\cite{frenkel83,magda}. The slope changes around $\rho=5.5$. Line
is for helping eyes } \label{fig:bz2}
\end{figure}

The equation of state i.e., dependence of $P$ on the $\rho$ is
shown in figure (2). In fact $P$ does not linearly increases with
$\rho$. The slope changes about the same value $\rho=5.5$ which
is in agreement with MC results of Frenkel and Eppenga
\cite{frenkel85}. This marks that from translational viewpoint
the system is totally distinct from ideal gas.

\begin{figure}
\centering
\includegraphics[width=7.5cm]{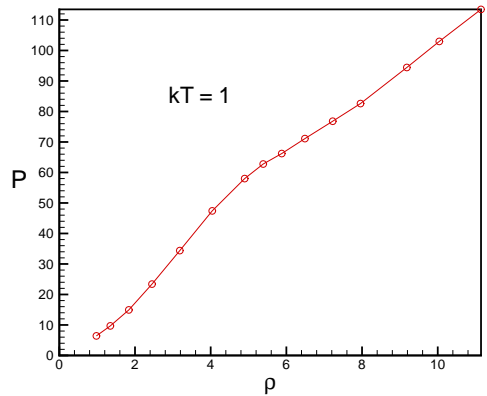}
\caption{ Fig.2: System pressure $P$ (reduced units) vs $\rho$.
Pressure slope changes around $\rho=5.5$. Line is for helping
eyes. } \label{fig:bz2}
\end{figure}

Next we turn to order parameter. The orientational nematic order
parameter $S$ is defined as follows:

\be S= \frac{1}{N^2} \langle \sum^N_{i,j=1}
\cos(2\theta_i-2\theta_j) \rangle. \ee

In which $\theta_i$ is the director angle of $i$-th needle with
positive $x$ axis and the average is taken over trajectories.
Dependence of $S$ on $\rho$ is sketched in figure (3).

\begin{figure}
\centering
\includegraphics[width=7.5cm]{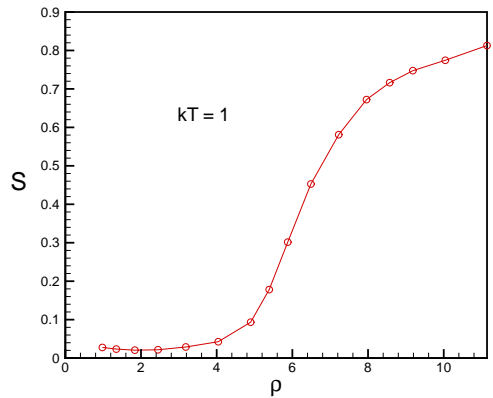}
\caption{ Fig.3: Dependence of nematic order parameter $S$ on
$\rho$. The transition to nematic ordered phase is expected to be
a finite size effect. Line is for helping eyes. } \label{fig:bz2}
\end{figure}

Our simulation shows a large value for the nematic order
parameter $S$ at high densities which corresponds to a nematic
phase. This result is expected to be a finite-size effect. We
argue that the possibility of truly phase transition to a nematic
phase is not excluded in 2D. More concisely, the conditions of the
Mermin-Wagner theorem are not satisfied here due to the fact that
inter-molecular potential between needles $V(r,\theta)$ is not
separable. Despite Monte Carlo simulations have not shown the
existence of such isotropic-nematic transition in a 2D gas of
needles \cite{frenkel85}, a quasi long range order of
Kosterlitz-Thouless type \cite{frenkel85,kosterlitz} was shown to
persist in the system. The subject of temporal velocity auto
correlation function has been the extensively studied in
\cite{frenkel83,magda} within the framework of MD and recently in
\cite{chrzanowska02,chrzanowska04} both in event-oriented MD
approach and analytically in the framework of Enskog kinetic
theory. We have explored the autocorrelation between longitudinal
and transverse decomposition of velocity. These quantities are
defined as follows:

\be
C_{||}(t)=\frac{1}{\langle v^2(0)\rangle}\langle \vec{v}(t).\hat{u}(0) \vec{v}(0).\hat{u}(0) \rangle.
\ee

\be
C_{\perp}(t)=\frac{1}{\langle v^2(0)\rangle}\langle \vec{v}(t)P\vec{v}(0) \rangle.
\ee

$\hat{u}$ denotes the unit vector along the needle orientation
and the matrix $P=1-\hat{u}(0)\hat{u}^t(0)$ is the projection
operator. In figure (4) we exhibit the temporal dependence of
$C_{||}(t)$ and $C_{\perp}(t)$ for various densities.

\begin{figure}
\centering
\includegraphics[width=7.5cm]{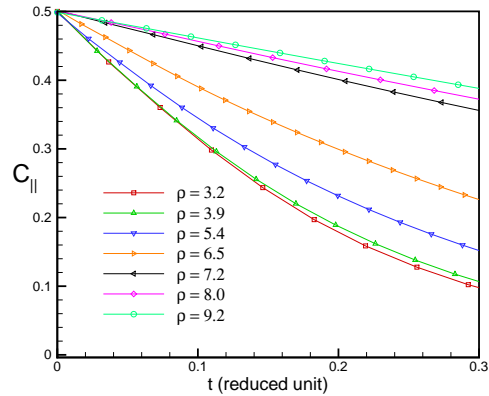}
\includegraphics[width=7.5cm]{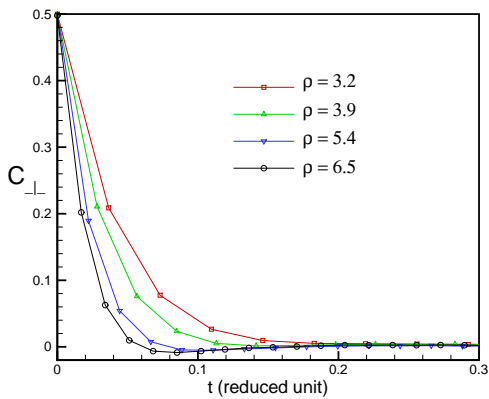}
\caption{ Fig.4: Temporal dependence of longitudinal and
transverse components of velocity ACF. Longitudinal component
exhibits a slower decay which is due to direction of impulsive
force between needles. } \label{fig:bz2}
\end{figure}

We refer the readers to \cite{chrzanowska02} for a detailed
discussion on velocity autocorrelations. We now consider the
temporal auto correlation of the second order angular order
parameter $C_2(t)$. This quantity is defined as follows:

\be
C_{2}(t)=\langle P_2(\hat{u}(t).\hat{u}(0)) \rangle.
\ee

In which $P_2$ is the second order Legendre polynomial. Figure (5)
shows the temporal dependence of $C_{2}(t)$.

\begin{figure}
\centering
\includegraphics[width=7.5cm]{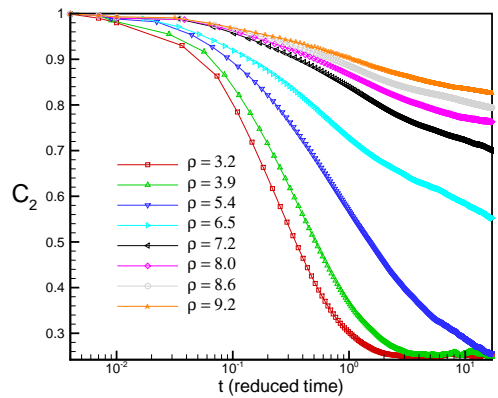}
\caption{ Fig.5: Temporal dependence of $C_2$ at various densities
(semi-log scale). } \label{fig:bz2}
\end{figure}

In low densities we see a fast decay which is attributed to the
fluid like behaviour. By increasing the density the temporal
behaviour becomes slow and two characteristics time scale emerge.
This confirms the nontrivial role played by the angular degree of
freedom. We emphasize that simulation on larger system is needed
to ensure the persistence of such large correlations. The
intermediate self scattering function $F_s(q,t)$ is shown in
figure (6).

\begin{figure}
\centering
\includegraphics[width=7.5cm]{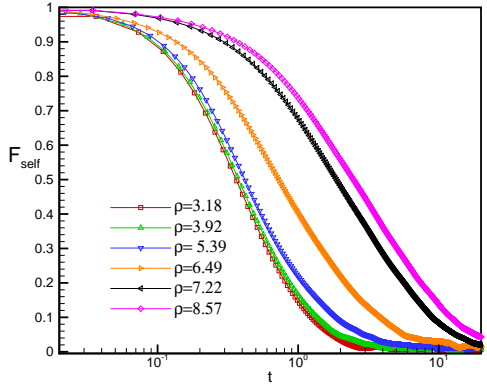}
\caption{ Fig.6: Time dependence of the intermediate self
scattering function  $F_s(q,t)$ at $q=3$. } \label{fig:bz2}
\end{figure}

\subsection{Structural properties}

We have obtained and explored various structural quantities. The
radial distribution function $g(r)$ (not shown) is featureless
for $r>1$ and regardless of the density value it approaches to
unity without showing any significant amplitude fluctuations.
This demonstrates that the gas posses no positional ordering.
However, the existence of excluded volume effect discriminates its
positional features to ideal gas. Next we consider the angular
spatial correlation function $g_2(r)$. This quantity is defined
as \cite{frenkel85}: $ g_{2}(r)=\langle
\cos(2[\theta(r)-\theta(0)]) \rangle. $ The average is over all
needle pairs having CM to CM distance $r$. Figure (7) plots the
dependence of $g_2(r)$ vs $r$.

\begin{figure}
\centering
\includegraphics[width=7.5cm]{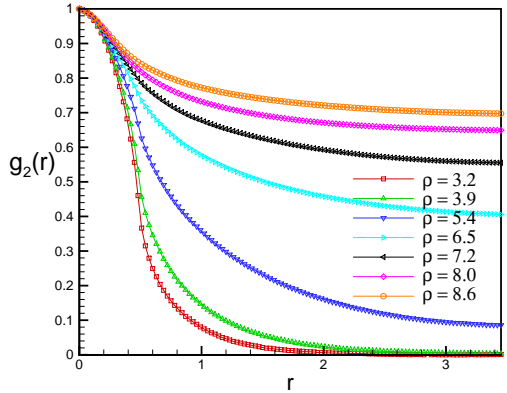}
\caption{ Fig.7: Dependence of $g_2(r)$ on $r$ for various
densities. } \label{fig:bz2}
\end{figure}

For small densities, $g_2(r)$ rapidly approaches zero. This
decrease is faster than algebraic. Contrary, for high density
$\rho >6$ we have a slow decrease which can be indicative of slow
dynamics and angular structural arrest. Taking the decay form
algebraic as $g_2(r) \sim r^{-\eta_2}$ we have found out by
fitting a curve the decay exponents at various densities. Figure
(8) shows our exponents which are compared to those of Frenkel and
Eppenga \cite{frenkel85}. At high densities MD and MC results are
in good agreement while quite notable differences are seen at low
densities.

\begin{figure}
\centering
\includegraphics[width=7.5cm]{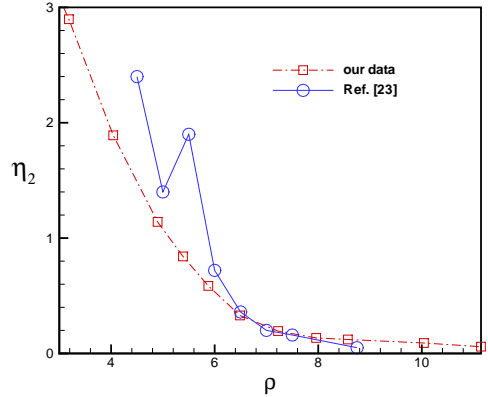}
\caption{ Fig.8: Dependence of algebraic decay exponent $\eta_2$
on $\rho$. } \label{fig:bz2}
\end{figure}

\section{Transport properties}

Now we report our results for the transport properties and
orientational structure of the system. We begin with
translational diffusion coefficient $D_{trs}$ defined as follows:

\be
D_{trs}=\frac{1}{2dNt}\langle \sum^N_{i=1} |\vec r_i(t) - \vec r_i(0)|^2\rangle
\ee

The term in the bracket is the translational mean-square
displacement (MSD) $ \langle (\Delta \vec{r})^2\rangle$. Here the
spatial dimension $d$ is two and the average is doubly taken over
trajectories of needles' CM and time origins. Figure (9) sketches
the time dependence of the translational MSD (see journal printed
version to see figure nine). After a ballistic regime, one
recovers normal diffusive behaviour. This confirms that from
translational viewpoint, the system resembles an structureless
gas. Figure (10) plots a CM trajectory of a typical needle at two
densities $\rho=3.9$ and $\rho=7.9$ :

\begin{figure}
\centering
\includegraphics[width=7.5cm]{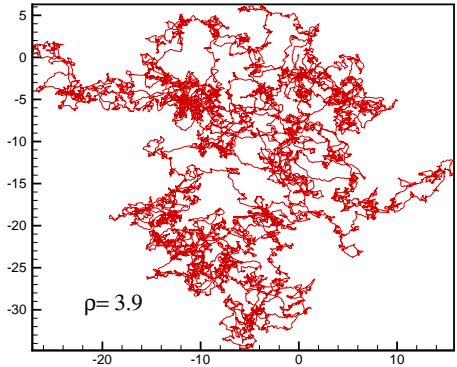}
\includegraphics[width=7.5cm]{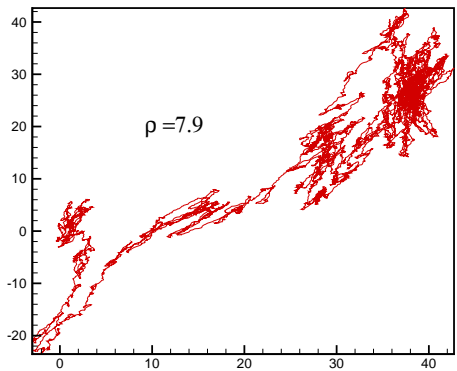}
\caption{ Fig.10: CM trajectory of a needle at $\rho=3.9$ during
1586 time unit and $\rho=7.9$ during 2329 time unit. Note the
unequal axes scaling. The true trajectories are elongated along
the axis having larger scale. } \label{fig:bz2}
\end{figure}

Three distinctive kinds of motion can be identified: diffusion,
channeling and entanglement among cages formed by neighbouring
needles. At low densities the motion resembles a normal
diffusion. At higher densities a channeling type of motion
emerges and at higher densities the topological constraints in 2D
leads to entanglement. Let us now explore the rotational
diffusion $D_{rot}$. In sharp contrast, $D_{rot}$ seems to be
entirely of different nature. We have:

\be
D_{rot}=\frac{1}{2\zeta Nt}\langle \sum^N_{i=1} |\theta_i(t) - \theta_i(0)|^2\rangle
\ee

Analogously the bracketed term denotes the angular mean square
displacement $\langle (\Delta \theta)^2\rangle$ and $\zeta$ is
the number of angular degrees of freedom (here $\zeta=1$). Figure
(11) sketches the time dependence of $\langle (\Delta
\theta)^2\rangle$ (see journal printed version to see figure 11).
A significant difference is seen compared to the translational
diffusion. For $\rho \geq 8$ the rotational MSD exhibits a
three-stage regime which can be attributed to angular glassy
dynamics. This is consistent to the density proposed in
\cite{frenkel85}. The possibility of angular glassy behaviour has
been earlier explored by Renner et al. \cite{renner}. It was
shown above a $\rho_c$, the angular auto correlations exhibit
slow dynamics and multi step relaxation which can be attributed
to glassy behaviour. Our findings is supportive of the existence
of angular glassy behaviour even when the translational degree of
freedom is released. We stress that simulations with a larger
system size is crucially needed to confirm this conclusion.
Finally we discuss the dependence of translational and rotational
diffusion constants on $\rho$ (figures 12 and 13).

\begin{figure}
\centering
\includegraphics[width=7.5cm]{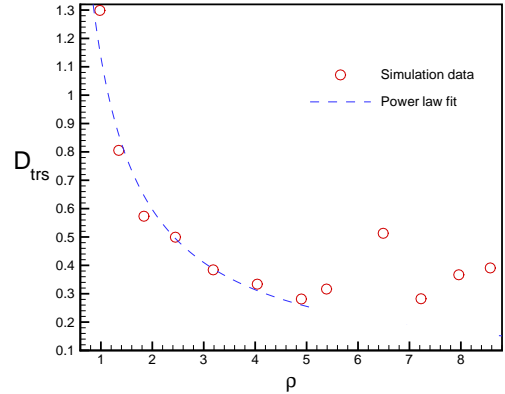}
\caption{ Fig.12: Dependence of $D_{trs}$ on needles density
$\rho$. } \label{fig:bz2}
\end{figure}

For $\rho<5$ translational diffusion constant $D_{trs}$ shows a
decreasing behaviour. This is in accordance with the Enskog
kinetic theory \cite{frenkel83}. In fact at low densities
successive binary collisions are uncorrelated therefore both
$D_{trs}$ and $D_{rot}$ would be inversely proportional to $\rho$
\cite{allen89}. For large densities $D_{trs}$ shows an increasing
trend which is in qualitative agreement with the 3D results
\cite{magda}. Frenkel and Maguire have exploited the Doi-Edwards
theory, valid above semi-dilute concentrations, and developed a
theory for the translational motion in 3D \cite{frenkel83}.
According to their results $C_{||}$ scales as $D_{rot}^{-0.5}$.
Combining this with the scaling result $D_{rot} \sim \rho^{-2}$
and that $D_{||}$ is the integral of $C_{||}(\tau)$ over time
they concluded that parallel component of translational diffusion
tensor scales as $D_{||} \sim \rho^{\frac{1}{2}}$. Moreover,
transverse component of diffusion tensor scales as $D_{\perp}
\sim \rho^{-\frac{1}{2}}$ within Doi-Edward theory. Therefore
they concluded that at high densities $D \sim
\rho^{\frac{1}{2}}$. The MD data of Magda et al. \cite{magda}
gives the dependence of $D_{\perp}$ on $\rho$ as $D_{\perp} \sim
\rho^{-1.57}$. Our results in figure (12) is in qualitative
agreement with the result obtained for a 3D gas of hard needles
\cite{frenkel83,magda,mukoyama1}. Note that
$D=\frac{1}{2}(D_{\perp} + D_{||})$ and at high densities it is
dominated by the longitudinal component. The fitted exponent to
the portion $\rho \leq 5$ of our result for translational
diffusion gives $D_{trs} \sim \rho^{-0.93}$. The value $0.93$
slightly differs to the predicted exponent $1$ by Enskog theory.
This shows dimensionality notably affects the system properties.
Eventually figure (13) exhibits the dependence of $D_{rot}$ on
$\rho$. In 3D based on scaling arguments it can be concluded that
within the Doi-Edward theory $D_{rot}$ scales as $\rho^{-2}$. MD
simulations for a 3D gas of hard needles gives the dependence of
$D_{rot}$ on density as $D_{rot} \sim \rho^{-\beta}$ with $\beta
\in [1.8,2.2]$ \cite{frenkel83}. The MD results of Magda et al.
\cite{magda} give the exponent $\beta=1.5$ for $32 \leq \rho L^3
\leq 72$ and $\beta=1.89$ for the density range $72 \leq \rho L^3
\leq 100$ ($L$ is the needle length). We have fitted both an
algebraic curve $D_{rot} \sim \rho^{-\beta}$ and an exponential
curve $D_{rot} \sim e^{-\frac{ \rho}{\xi}}$ to our own data. It
turned out that $\beta=3.6$ and $\xi=0.92$ with $\chi^2=0.7828$
and $0.9375$ for the power law and exponential fits
correspondingly (exponential curve is better fitted). We thus
observe a slower decrease for $D_{rot}$ in 2D rather than in
three. This can be explained on the basis of having more
pronounced degree of entanglement and topological constraints
among needles in two dimensions respect to 3D.

\begin{figure}
\centering
\includegraphics[width=7.5cm]{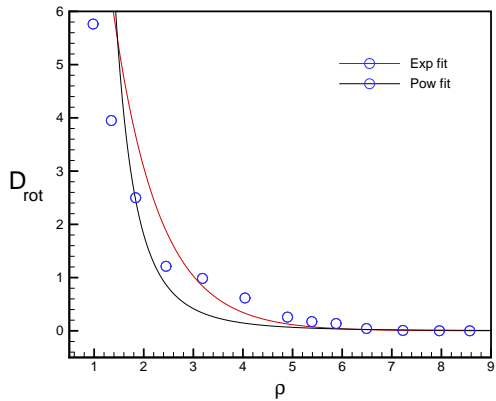}
\caption{ Fig.13: Dependence of $D_{rot}$ on $\rho$. }
\label{fig:bz2}
\end{figure}

\section{Summary and Concluding Remarks}

We have simulated the dynamics of a 2D gas of hard needles by
event-oriented molecular dynamics. Many of the temporal
autocorrelation functions both translational and angular exhibit
a sort of slow dynamics and multi step relaxation. The most
interesting feature is the existence of three regimes in the
angular mean squared displacement. This can be attributed to slow
dynamics. Our findings show relaxing the translational degrees of
freedom does not smear out angular slow dynamics. Density
dependence of translational and rotational diffusion coefficients
has been obtained and compared to three dimensional results. In
2D dependence of the translational diffusion coefficient on
density qualitatively resembles to 3D. Rotational diffusion
constant exhibits an algebraic decay but with a larger exponent
than in 3D.

\section{acknowledgement}

This work has been funded by {\it Iran National Science
Foundation } under the grant number 844169. We would like to
deeply express our gratitude to Mainz university for the kind
hospitality and computing facility during our visit to Professor
Kurt Binder group where parts of this project were carried out.
We are highly indebted to Tanja Schilling for very stimulating
and fruitful discussions and enlightening comments. Enlightening
comments of an anonymous referee is appreciated. Useful
discussions with Mir Faez Miri is appreciated. MEF is thankful to
B. Vazirol Vozaraa for useful help.

\bibliographystyle{unsrt}

\end{document}